\documentclass{article}
\usepackage[utf8]{inputenc}
\usepackage{xcolor}
\usepackage{authblk}

\title{Astrochemistry associated with planet formation\footnote{to appear as a book chapter in {\it ExoFrontiers: big questions in exoplanetary science}, Ed. N.\ Madhusudhan (Bristol: IOP Publishing Ltd.), AAS-IOP ebooks, authors' version} }
\author[1]{Ewine F. van Dishoeck}
\author[2]{Edwin A. Bergin}
\affil[1]{Leiden Observatory, Leiden University, the Netherlands}
\affil[2]{Astronomy Department, University of Michigan, USA}
\date{}
\usepackage{natbib}
\usepackage{graphicx}
\usepackage{hyperref}
\usepackage{geometry}
\usepackage[utf8]{inputenc}

\newcommand {\apj} {{\it ApJ}}
\newcommand {\apjs} {{\it ApJS}}
\newcommand {\apjl} {{\it ApJL}}
\newcommand {\aap} {{\it A\&A}}

\newcommand {\aapr} {{\it A\&A Review}}
\newcommand {\araa} {{\it ARA\&A}}

\newcommand {\mnras} {{\it MNRAS}}

\newcommand {\nat} {{\it Nature}}

\newcommand {\gca} {{\it Geo. Cosmo. Acta}}

\newcommand {\icarus} {{\it Icarus}}
\newcommand {\ssr} {{\it Space Sci. Rev.}}

\begin{document}
\maketitle

\section{Introduction} 

Stars and planets are formed deep inside dense clouds. When these
clouds collapse, the surrounding gas and dust become part of the
infalling envelopes and rotating disks, thus providing the basic
material from which new planetary systems are made.  Today we are
finding that much of the chemical composition of the planet-building
material is likely set in the cold pre- and protostellar stages and
preserved en route to planet and comet construction sites. With new
observational techniques, astronomers can now zoom into these
planet-forming disks and study them on scales comparable with the
orbit of Saturn in our own solar system \citep{Andrews20}.

Astrochemistry, also known as molecular astrophysics, is the study of
the formation, destruction and excitation of molecules in astronomical
environments.  By observing many lines of the same species, molecules
are excellent diagnostics of the physical conditions and their high
resolution spectral profiles provide unique information on the
kinematics in the regions where they reside. More than 200 different
molecules have now been detected in interstellar space
\citep{McGuire18}. The main questions in astrochemistry therefore
include: how, when and where are these molecules produced? What do
they tell us about temperatures, densities, gas masses, ionization
rates, radiation fields, and dynamics of the star-forming clouds and
planet-forming disks? From these quantities what can be learned about
the physics of star and planetary birth? How are they cycled through
the various phases of stellar evolution, from birth to death? How far
does chemical complexity go? And, most far-reaching, can they form the
building blocks for life elsewhere in the Universe?  While organics
are a central focus of the latter question, water, in the form of ice,
stands out as a reservoir of abundant elemental oxygen throughout all
phases, hinting that life's solvent started its journey in space
\citep{vanDishoeck14ppvi}.

This brief overview summarizes some of the main results and questions
regarding the chemistry along this journey from cloud to
disk. References are limited primarily to review papers of the many
hundreds of papers in this rapidly growing field; citations to
individual papers are mostly to relevant results in the last few years
and are highly incomplete. Readers are encouraged to cite original
papers.

\begin{figure}[t]
\begin{center}
\includegraphics[width=0.7\textwidth]{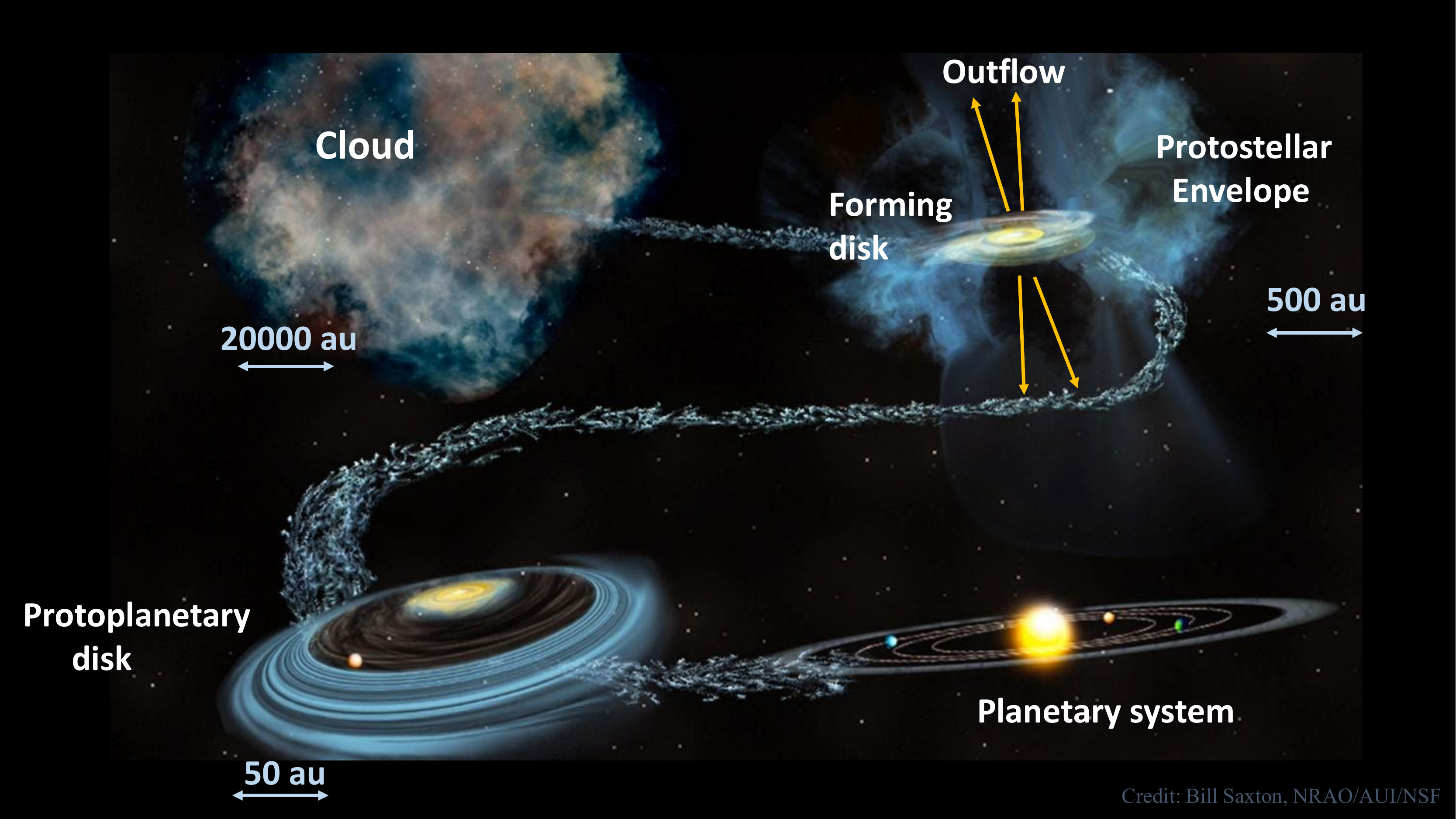}
\end{center}
\caption{Stages in the formation of a new planetary system from a dark interstellar cloud. Much of the chemistry of planet-forming material is likely set in tihe cold pre- and protostellar stages. Also, there is strong evidence that the first steps of planet(esimal) formation take place early, in the embedded protostellar phase of star formation (typically few $\times 10^5$ yr after collapse). Artist impression from Bill Saxton (NRAO/AUI/NSF), annotated by the authors.}
\label{fig:saxton}
\end{figure}

\section{State of the art}  

Recent overviews of the field of astrochemistry have been given by
\cite[][and refs cited]{Tielens13,Yamamoto17}. Fourteen challenges
have been formulated by \cite{vanDishoeck18}.

\subsection{How to observe molecules}

Molecules can be observed through their electronic, vibrational or
rotational transitions at optical/UV, infrared and millimeter
wavelengths, respectively. Several of the most powerful new telescopes
deployed over the past decades have been particularly well suited to
observe interstellar molecules. The bulk of the molecules are detected
through their rotational transitions at millimeter wavelengths, and
various single-dish and interferometers at millimeter wavelengths with
increasingly sensitive broad-band detectors have become available
culminating in the Atacama Large Millimeter/submillimeter Array
(ALMA). ALMA has the combined sensitivity and spatial resolution to
zoom in on planet-forming disks. Millimeter observations require the
molecule to have a permanent dipole moment to be detectable.

Infrared observations are highly complementary. Space missions like
the {\it Infrared Space Observatory}, {\it Spitzer Space Telescope}
and the {\it Herschel Space Observatory} at mid- and far-infrared
wavelengths were particularly well suited to study H$_2$O and
molecules like CO$_2$ that are abundant in our Earth's
atmosphere. Moreover, solid state species (silicates, ices), complex
molecules (PAHs, fullerenes) and small molecules without a dipole
moment like CH$_4$, C$_2$H$_2$ and CO$_2$ can be uniquely observed
through their vibrational transitions. Warm H$_2$ can also be detected
through mid-infrared lines, but cold H$_2$ in dark clouds is invisible
and can only be traced indirectly through other molecules, most
notably through far-infrared lines of HD.  These space-based data have
been complemented by studies of selected molecules using ground-based
8m optical/infrared telescopes equipped with high resolution
spectrometers. The {\it James Webb Space Telescope} (JWST), to be
launched late 2021, will be the next big jump in mid-infrared
capabilities.

\subsection{Astrochemical models}

Chemistry in the interstellar medium, where densities are typically
10$^3 - 10^6$ cm$^{-3}$, cannot be described by the laws of
thermodynamics, except in the densest stellar and planetary
atmospheres well shielded from radiation, with densities $> 10^{13}$
cm$^{-3}$. In interstellar space, the chemistry is controlled by
two-body reactions and abundances can be obtained through kinetics
involving large networks of reactions. These networks contain
gas-phase reactions (ion-molecule, neutral-neutral) as well as
gas-grain and grain-surface chemistry, with UV photoprocesses playing
a role in both gaseous and solid phases. Thanks to many decades of
laboratory experiments and quantum chemical calculations by physicists
and chemists, deep insight into these processes has been obtained and
their reaction rates under different conditions have been quantified.

Compilations of the rate coefficients together with codes that solve
the coupled differential equations of these networks include the \href{http://udfa.ajmarkwick.net/index.php?mode=downloads}{UMIST
2013 database}, \href{http://kida.astrophy.u-bordeaux.fr}{the KIDA database}, and \href{https://uclchem.github.io}{the UCLCHEM code}. Programs
tailored to exoplanetary atmospheres include \href{https://github.com/exoclime/VULCAN}{VULCAN} and \href{https://arxiv.org/abs/1905.06826}{LEVI}.

To run an astrochemical model, a prescription of the temperature and
density structure of the source, the incident UV radiation field, the
cosmic ray ionization rate, as well as the abundances of the
(volatile) elements that are not locked up in refractory solids but
that can cycle between the gas and ice phases are needed. Given the
uncertainties in each of these parameters as well as in individual
rate coefficients, agreement between observations and models at the
factor of a few level is generally considered to be good.

\section{Important questions + goals: chemistry from clouds to planets} 

The various stages in the formation of a new planetary system are
illustrated in Fig.~\ref{fig:saxton}. The first three stages typically
take a few million years, with a clear drop in gas and dust content
after $\sim$3 Myr. A key recent finding is that planet formation must
already start early, in the embedded phase of star formation a few
$\times 10^5$ yr after cloud collapse, in order to account for the observed
mass in exoplanet cores \citep{Tychoniec20}.  This is consistent with
the solar system record as Fe-rich meteorites sample the cores of
large bodies that formed, in many instances, within the first few
100,000 years \citep{Kruijer17}. Thus, much of the chemistry of
planet-forming material is already set in the cold pre- and
protostellar stages. On the other hand, there is clear evidence that
molecules and dust were vaporized in the inner young solar nebula and
then re-condensed, based on elemental abundance patterns and minerals
found in meteorites \citep{Connelly12}. Key questions therefore
include

\begin{itemize}

\item What determines the chemical composition of
planet-building material: inheritance or full reset? At what disk
radius is the cross over (for a given type of star)?  Can the composition of the young still-forming protostellar disks be characterized enough to reveal answers to these questions?

\item If inherited from the collapsing cloud, to what extent is the
  gas and ice composition modified en route from cloud to disk (e.g.,
  by accretion shocks), and within the protoplanetary disk midplane?
  
\item Ice rich pebbles drift inwards during the main phase of
  planetary assembly.  What is the fate of pebbles drifting from the
  outer tens to hundreds of au into the inner disk ($<$10 au)?  How
  does this impact planet formation and its chemistry?
 
\item How quickly are ices locked up in large planetesimals, and what
  are the C/O or C/N ratios with radius in the midplane?  How are
  these affected by pebble drift and dust traps? How do the snowlines
  move with time?  How important is vertical mixing?

\item Are the majority of the heavy volatile elements in a giant
  planetary atmosphere accreted as ice or as gas? How are they
  modified in the atmosphere itself and does a dilute core mix with
  the envelope?  What is the importance of LTE versus kinetic
  chemistry in the outer layers?

\item What can the composition of exo-planetary atmospheres ultimately
tell us about their formation location, given the above questions?   

\item What can we learn about terrestrial planet formation and the delivery of needed volatile elements from astronomical observations?

\end{itemize}

\subsection{Cold dark clouds}

Chemistry starts in cold dark clouds prior to star formation with have
typical temperatures of 10 K and densities of 10$^4$-10$^5$
cm$^{-3}$. They exhibit a variety of chemical characteristics, the
most important of which are ice formation and heavy deuterium
fractionation \citep{Bergin07,Caselli12aar,Ceccarelli14}. Most of these
chemical signatures are transferred to the protostellar stage where
they are observed in the cold outer parts of the collapsing envelope
(Fig.~\ref{fig:saxton}).

Infrared observations of ice features toward reddened background stars
show that the formation of water ice starts once the cloud extinction
A$_V$ is a few mag \citep{Boogert15,Oberg11}. The formation of water ice on
the surfaces of grains is now well characterized in the laboratory and
models, and tested against observations of cold water gas and ice
\citep{vanDishoeck13,vanDishoeck14ppvi}. CH$_4$, NH$_3$ and some CO$_2$
ice are also made in this early phase, with the amount of ice rapidly
increasing as the cores become more centrally concentrated. All of
these molecules are made on the grain surfaces, from reactions of
atomic O, C and N with atomic hydrogen; they are not accreted from the
gas.

At densities around 10$^5$ cm$^{-3}$, the timescales for freeze-out
become shorter than the lifetime of the cloud core, and CO (the
dominant form of volatile carbon at these high densities) rapidly
depletes from the gas onto the grains. This CO-rich ice can
subsequently react with atomic H to form H$_2$CO and CH$_3$OH, a
process that has been demonstrated to proceed rapidly at low
temperatures in the laboratory. Various other routes can transform CO
to CO$_2$ and more complex organic ices, even at very low
temperatures.

A clear signature of cold interstellar chemistry are high abundances
of deuterated molecules such as DCN, with DCN/HCN ratios at least
three orders of magnitude higher than the overall [D]/[H] ratio of
$2\times 10^{-5}$. Even doubly- and triply-deuterated molecules such
as D$_2$CO and ND$_3$ have been detected. This huge fractionation has
its origin in two factors. First, the zero-point vibrational energy of
deuterated molecules is lower than that of their normal counterparts
because of their higher reduced mass. This makes their production
reactions exothermic. In cold cores, most of the fractionation is
initiated by the H$_3^+$ + HD $\leftrightarrow$ H$_2$D$^+$ + H$_2$
reaction which is exothermic by about 230 K. H$_2$D$^+$ then transfers
a deuteron to CO or N$_2$ or another species. Since CO is the main
destroyer of both H$_3^+$ and H$_2$D$^+$, their abundances are even
further enhanced when CO is removed from the gas. Similarly,
CH$_2$D$^+$ is enhanced at low temperatures and can enhance deuterium
in organic molecules. The second, related effect is that the gaseous
atomic D/H ratio is enhanced, which implies that relatively more D
than H arrives on the grain to react with CO to make deuterated
versions of formaldehyde, methanol and more complex organic molecules,
and powering deuteration of water ice.

Key references summarizing this stage are \cite{Caselli12aar,Boogert15}.

\begin{figure}[t]
\begin{center}
\includegraphics[width=0.75\textwidth]{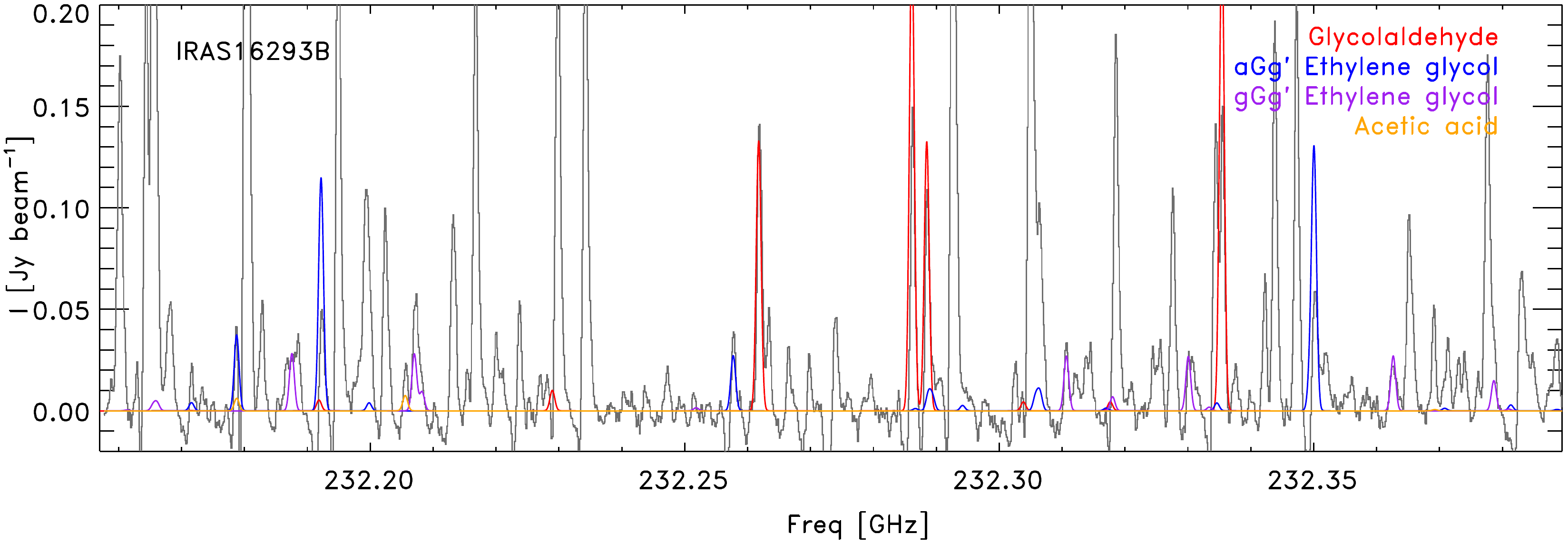}
\end{center}
\vspace{-0.8cm}
\caption{Blow-up of a small part of the ALMA spectral survey toward
  the young solar-mass protostar IRAS16293B on scales of 50 au,
  illustrating the wealth of lines (roughly 1 line every 3 km
  s$^{-1}$) and the identification of complex organic molecules like
  glycolaldehyde and ethylene glycol \citep{Jorgensen16}.}
\label{fig:iras16293}
\end{figure}

\subsection{Protostellar envelopes and young disks}

Once a protostar has formed in the center, its luminosity sets up a
temperature gradient in the gas and dust. Temperatures increase from
10 K in the outer envelope to a few hundred K in the innermost region
close to the protostar. This can result in numerous chemical changes:
radicals become mobile in and on icy surfaces and recombine to form
even more complex molecules, whereas increased UV and X-rays from the
protostar-disk accretion boundary trigger further chemistry in the ice
and gas.  When dust temperatures become high enough for ices to
sublimate, molecules do so presumably in a sequence according to their
binding energies, with volatile species like CO and N$_2$ sublimating
first.

Once dust temperatures of $>100$ K are reached, even the
strongly-bound water and methanol ice sublimate, together with any
minor molecules trapped in them, resulting in particularly rich
gas-phase millimeter spectra (Fig.~\ref{fig:iras16293}). This inner 100
K zone is called the `hot core'. Here high-temperature gas-phase
reactions between sublimated molecules can result in `second
generation' complex organic molecules \citep{Balucani15}.

Much of the extensive chemical complexity seen in hot cores is thought
to be largely assembled in ices.  A major question is whether this
rich chemistry is incorporated in the forming disks
(Fig.~\ref{fig:saxton}), and if so in what form. Further, the full
extent of chemical complexity occurring on grain surfaces is still
uncertain.  It is clear that some molecules of `pre-biotic'
significance are found in ices, such as simple sugars and peptide
bonds (Fig.~\ref{fig:iras16293}), but these are still small compared
with molecules found in biological systems as well as the
macromolecular material found in meteorites and comets (\S 3.5).

The solar system record suggests that there must have been a mixture
of material where the chemistry is reset with pre-existing (or less
processed) material `inherited' from the interstellar medium (\S 3.5).
In this regard, it is unclear whether the material entering the disk
experiences a strong accretion shock, and if so, whether this shock
completely resets the chemistry or whether strongly bound ices
survive. At face value observations find an interesting transition in the disk
seen in hydrocarbon emission and SO, but in general the overall
temperature of these systems appears to be rather cool ($< 50$~K).

Key references for this section include
\cite{Herbst09,Ceccarelli14,Jorgensen20}.

\begin{figure}[t]
\begin{center}
\includegraphics[width=0.9\textwidth]{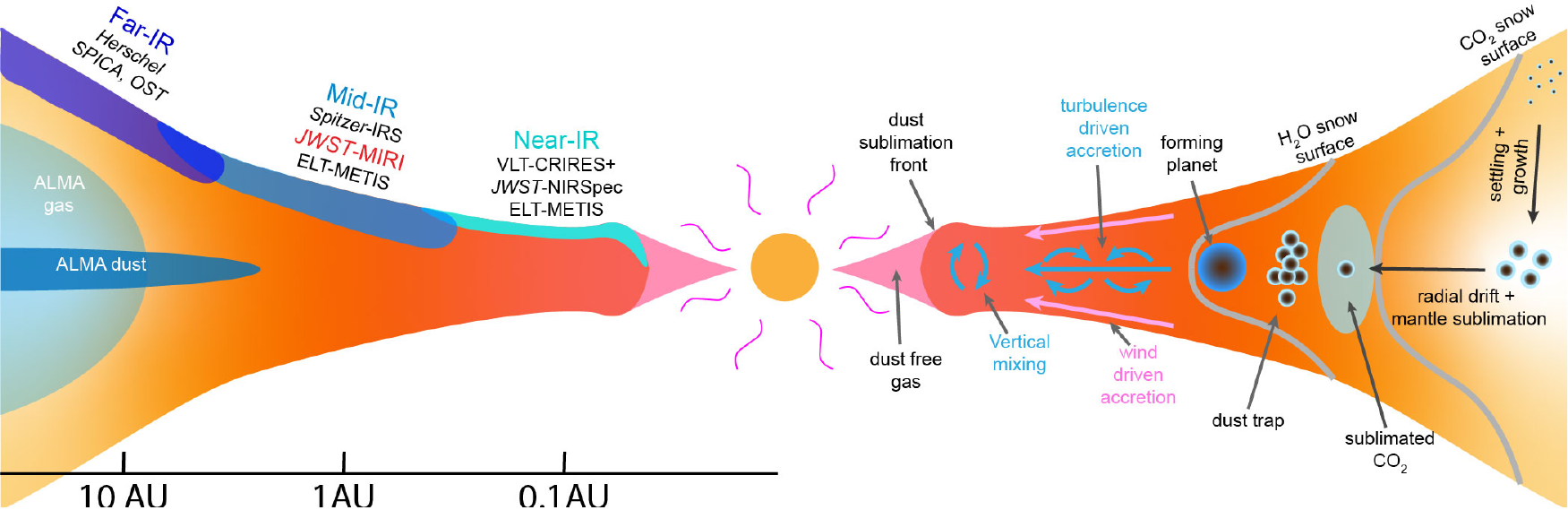}
\end{center}
\caption{Overview of the structure of planet-forming disks. The
  left-hand side shows the different parts of the inner disk probed by
  various instruments. The right-hand part illustrates some important
  physical and chemical processes, including gas accretion and mixing,
  as well as grain growth, settling and radial drift of icy pebbles,
  which sublimate at snow lines. Figure by A.\ Bosman, PhD thesis,
  Leiden University.}
\label{fig:bosman}
\end{figure}

\subsection{Planet-forming disks}

ALMA has been transformational for studies of planet-forming disks,
and will likely remain so for the coming decade(s)
\citep{Andrews20}. ALMA and high contrast infrared imaging of disks
are providing striking pictures of the first steps of planet
formation: rings, gaps, cavities, asymmetric structures, and / or
spiral arms. Although there are many possible interpretations of these
structures, they do indicate that growth of dust grains to pebbles and
planetesimals is taking place, starting even in the embedded phase.

Disks remain challenging to study both observationally and
theoretically, however. Observationally, molecular lines are very weak
since disks are small (typically less than 1$''$ on the sky) and their
mass is only 1\% of that of the collapsing cloud. Theoretically, they
are a challenge since they cover a huge range of densities and
temperatures in at least two dimensions, from $>1000$~K in the inner
disk and upper layers, to 10~K in the outer midplane, and from
densities of $> 10^{13}$~cm$^{-3}$ in the inner midplane down to
10$^5$~cm$^{-3}$ in the upper outer layers. UV radiation fields from
the central star impinging on the surface layers can be as high as
10$^5$ times the interstellar radiation field at 10 au from the star,
thereby ionizing atoms and dissociating molecules. Thus, different
types of chemistry are important in different parts of the disks
\citep{Henning13,Bergin18,Oberg20} (Fig.~\ref{fig:bosman}).

Moreover, gas and dust are largely decoupled (except for the smallest
grains): dust grains grow to pebble size (few cm), settle to the
midplane and drift in radially (Fig.~\ref{fig:bosman}). If they
encounter a pressure bump, dust traps can form where particles can
grow to even larger, planetesimal sizes. Gas/dust ratios can therefore
differ significantly from 100. While ALMA is particularly well suited
to study the gas and mm-sized dust in the outer disk ($>10$ au for
gas, few au for dust), JWST and other mid-infrared facilities probe
the important inner 10 au planet-forming zones of disks.

The decreasing temperature in the radial direction sets up a range of
snowlines, i.e., radii where molecules freeze-out onto the grains,
defined as the half-gas, half-ice point. Because of the vertical
temperature gradient, the 2D snow surfaces are actually curved
(Fig.~\ref{fig:bosman}). Snowlines are thought to play a significant
role in planet formation, since ice coating of grains enhances the
solid mass, potentially promotes coagulation of grains to larger
particles and/or require a higher collision velocity for destruction
\citep{Blum18,Pinilla17}, effects which are particularly prominent
just outside the snowline.  Thus it is thought that water ice coated
grains grow to larger sizes and provide larger pebbles for planet
formation which has implications on growth via pebble accretion
\citep{Morbidelli15}. They also control the bulk elemental composition
of the icy planetesimals and gas from which exoplanetary atmospheres
are built.

\begin{figure}[t]
\begin{center}
\includegraphics[width=0.42\textwidth]{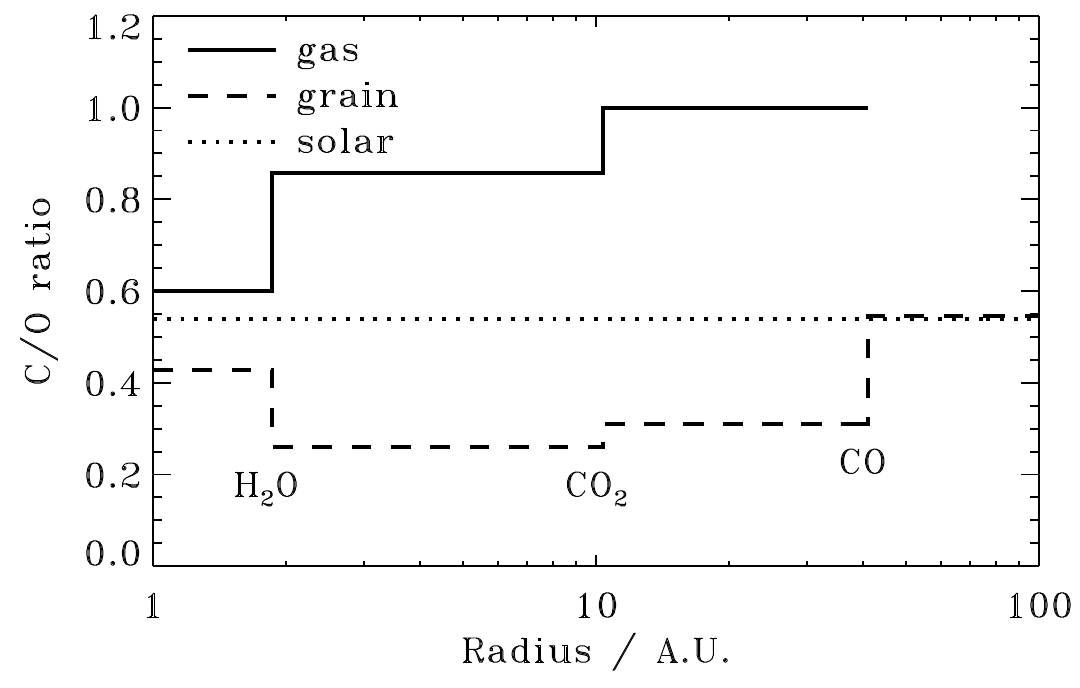}
\includegraphics[width=0.47\textwidth]{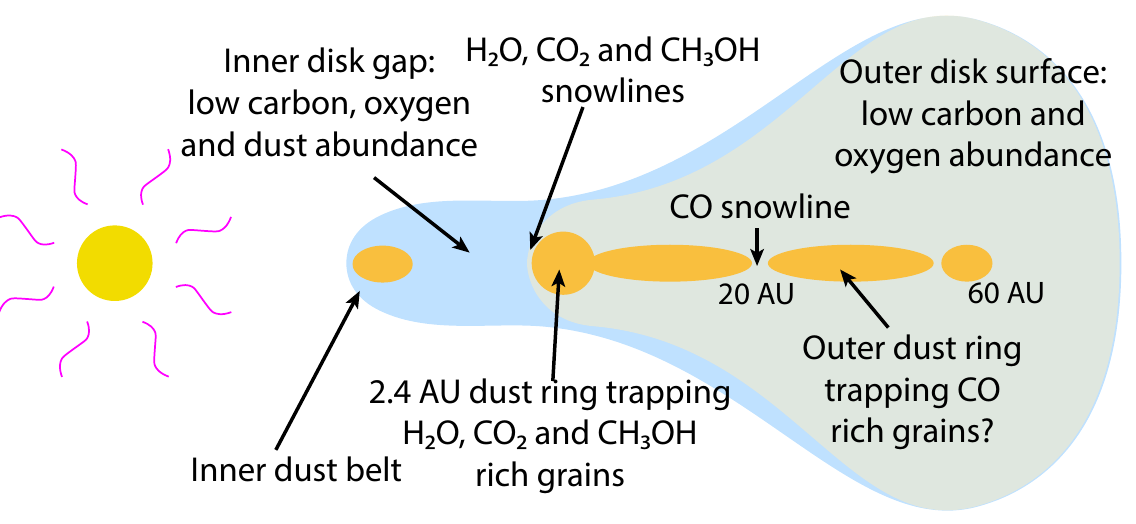}
\end{center}
\caption{Left: Midplane C/O ratios in gas and ice as function of disk
  radius, with major steps occurring at the H$_2$O, CO$_2$ and CO
  snowlines. The ice is oxygen rich, whereas the gas has high C/O but
  is overall depleted in carbon and oxygen \citep{Oberg11co}. Right:
  illustration how abundances in the inner disk can be affected by
  dust traps locking up volatiles in the outer disk beyond their snowlines,
  as in this case for the TW Hya disk \citep{Bosman19twhya}.}
\label{fig:oberg}
\end{figure}

{\bf Cold outer disks}. Most molecules detected in disks are simple
species containing only a few atoms. Only two complex molecules,
CH$_3$CN and CH$_3$OH, have been detected so far, and only
barely. Even the brightest disks do not show rich line forests such as
found for hot cores. This lack of gas-phase lines is because the bulk
of the disks are cold with molecules frozen out.  In fact, the CO snow
surface is resolved in highly flared systems - thus the comet-forming
zone can essentially be seen by eye in some ALMA images.  The most
important snowline, that of water, is generally out of reach because
it typically occurs at a few au for T Tauri stars, too small for ALMA
to pick up. Only systems undergoing luminosity outbursts such as V883
Ori or young disks with enhanced accretion luminosity, for which the
water snowlines have moved out to tens of au, offer the opportunity
for direct detection and imaging of complex chemistry \citep{Lee19}.

The overall picture of the outer disk is therefore that much of the
volatile oxygen and carbon is locked up in ices in large bodies.
Since the dominant H$_2$O and CO$_2$ ices have more oxygen than
carbon, the ice is oxygen rich, with an overall C/O ratio that is
lower than the interstellar or solar abundance (Fig.~\ref{fig:oberg})
\citep{Oberg11co}. In contrast, the gas is carbon rich even though it is
overall depleted in carbon and oxygen. Once C/O$>1$, the chemistry
completely changes and small hydrocarbon molecules become
abundant. This can indeed explain the strong observed C$_2$H and
c-C$_3$H$_2$ emission in some disks, although models suggest that midplane
chemistry tends to evolve toward gaseous C/O significantly less than 1
depending on the level of ionization \citep{Eistrup18}.  JWST will be
able to observationally constrain the ice composition in
planet-forming zones, but only for a handful of near edge-on disks and
then only in the intermediate layers.

{\bf Warm inner disk.} The warm gas in the upper layers of the inner
disk ($<10$ au) emits strongly at mid-infrared wavelengths. Indeed, a
dense forest of lines due to simple molecules has been detected by
{\it Spitzer} and ground-based infrared telescopes, providing a
glimpse of the chemistry in that region \citep{Pontoppidan14}.  Here
the high temperatures can completely `reset' the chemistry. Key
abundant molecules without a dipole moment such as CO$_2$ and
C$_2$H$_2$ are observed, together with H$_2$O, OH and HCN, allowing
tests of major C, O and N reservoirs and high temperature
chemistry. The degree to which these data can probe the disk midplane
chemistry and ice sublimation, relevant for planet formation models,
is still unclear however, since they can only be observed if effective
vertical mixing takes place. On the other hand, the recent
observational evidence for meridional flows suggests that giant
planets accrete a significant fraction of their gas from the upper
disk layers \citep{Teague19}. JWST will be poised to provide much
deeper searches for CH$_4$, NH$_3$ and minor species, and detect
isotopologues to constrain line optical depth, and follow the inner
disk chemistry (at least in the surface layers) from the youngest
embedded disks to the debris disk stage, and across the stellar type
range.

Key references for this stage include
\cite{Henning13,Pontoppidan14,Andrews20,Oberg20}.

\subsection{Disk evolution and planet(esimal) formation}

Disk structure and chemistry is not static but evolving, due to grain
growth and inward drift of dust particles (Fig.~\ref{fig:bosman})
\citep{Pinilla17}. At the same time, mass and angular momentum
transport in disks, both inward and outward, are still poorly
understood.  How does this affect the chemistry?  Recent ALMA data
indicate either surprisingly low gas/dust ratios or a large fraction
(up to a factor of 100) of volatile carbon, carried by CO, missing
from the gas phase.  The latter case, that CO is missing from layers
with temperatures $>$20~K (the CO sublimation temperature), has
support in terms of consistency with accretion rates, the mass needed
to make planetary systems, and, in a handful of instances, detections
of HD to constrain the gas mass \citep{Bergin18b}.  Thus, at face
value, it appears that planet formation removes volatiles from the
gas.

The first clue actually came from {\it Herschel}-HIFI observations
which revealed surprisingly weak gaseous H$_2$O lines in disks. Since
cold H$_2$O vapor in the outer disks results from UV photodesorption
of water ice coating small grains in intermediate layers, the observed
weak emission indicates that this water ice is spatially confined to
only a part of the disk: radially, vertically or both. The most
plausible interpretation is that most of the water is locked up in
large icy bodies (at least pebble size) in the midplane
\citep{Du17}. A second clue comes from models of dust coagulation and
growth which suggest that water and CO on the surface can be caught up
in the grain evolution and locked within icy pebbles and planetesimals
\citep{Krijt20}.  In the case of CO it may be combined with an active
disk chemistry transforming CO into CH$_3$OH, CO$_2$ or hydrocarbons.

Do these heavy elements return to the gas when the pebbles drift
inward and cross the snowlines (Fig.~\ref{fig:bosman},
\ref{fig:oberg})? If so, one would expect inner disk abundances inside
the water snowline to be similar to, or even higher than, the stellar
abundances. This is not always the case: there is significant evidence
that dust traps located beyond the major snowlines in the outer disk
can lock up large fractions of carbon and oxygen preventing these
pebbles to drift inward and resulting in inner disks that are depleted
in these elements (Fig.~\ref{fig:oberg}, right)
\citep{Kama15,McClure20}.

So how do we then determine the disk chemical composition just prior
to planet formation? Much of the chemistry of planet formation is
unfortunately hidden from our view, including the bulk C- and
O-containing species. One option is to observe warmer, younger disks
where less of the material is frozen out and fewer planetesimals have
formed, such as the outbursting case described above. Observational
studies show that only the youngest, deeply embedded sources have high
enough temperatures to reveal the full chemical richness (\S 3.2)
\citep{vantHoff20}. The forming disks around Class 0 sources, like
those for IRAS16293-2422 (Fig.~\ref{fig:iras16293}), may therefore
well provide the most detailed information of chemistry on
solar-system scales, even if the material has not yet settled into a
Keplerian structure. That chemistry, in turn, may have been set
already to a large degree in the dense cold core just prior to and
during collapse, arguing for an `inheritance' rather than a `reset'
scenario, at least in the outer disk.

Key references for this section include
\cite{Pinilla17,Du17,Bergin18b,Krijt20}.
 
\subsection{Clues from our solar system: comets and meteorites}

The solar system record shows evidence of both a full reset of the
chemistry as well as inheritance from the interstellar medium. The
pattern of elemental abundances in the Earth's lithosphere and in
meteorites suggests that this material formed within a hot ($>$1500~K)
atomic gas in which all original interstellar material was vaporized,
followed by a sequence of mineral condensation as the gas cooled
\citep{McDonough95}.  This cannot be the case for all solar system
materials, however, as the deuterium enrichment in Earth's water, and indeed all solar system water,
implies a cold ($<$30~K) source and cometary abundances appear to
follow that of the interstellar medium \citep{Altwegg19}.

Comets and other icy bodies in our own solar system provide the best
clues to the chemistry in the cold part of the natal solar nebula disk
and the level of change since their formation 4.5 billion years
ago. Comparisons between cometary and interstellar abundances,
including those from the Rosetta mission to comet 67P/C-G, have shown
some tantalizing similarities \citep{Mumma11,Drozdovskaya19}.  The
organic complexity in comets goes further than has been detected so
far in young disks, with even 4- or 5-carbon organic molecules,
alkanes, aromatics and even the simplest amino acid, glycine, detected
\citep{Altwegg19}. More bright comets like Hale-Bopp that can be
probed with ALMA and modern infrared instruments will provide further
insight into their origin, including through determining their D/H
ratio in water.

Primitive carbonaceous meteorites also have a rich organic
composition, but much of that inventory is either macromolecular or
altered via acqueous reactions in liquid water \citep{Alexander07}.
Thus, in terms of the origins of biology and the importance of
molecules produced via cold chemistry in the interstellar medium, a
key question is the amount of material that was supplied to the Earth
that originated in the asteroid belt or beyond. The fate of this
material on the Earth itself is, of course, a central issue (see \S
3.6).

Key references for this section include
\cite{McDonough95,Altwegg19,Alexander07}.

\subsection{Exoplanets and their atmospheres}

\begin{figure}[t]
\begin{center}
\includegraphics[width=0.8\textwidth]{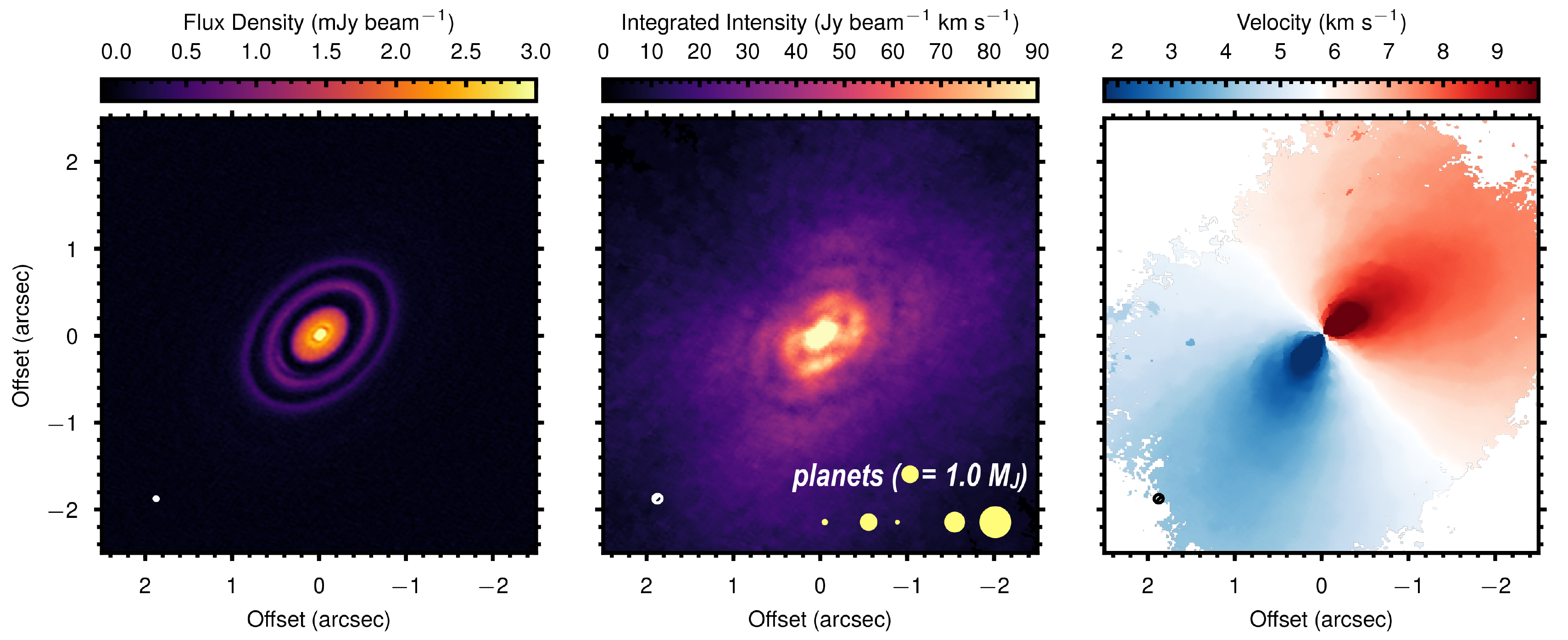}
\end{center}
\vspace{-0.7cm}
\caption{High spatial resolution ALMA image of the 1.3 mm
  dust emission from the HD~163296 disk ($d$=101 pc) showing several
  dust rings (left) and $^{12}$CO $J$=2--1 emission (middle) published
  by \cite{Isella18}. The right-hand panel is the H$_2$ gas velocity
  field in $^{12}$CO 2--1 taken from \cite{Teague19}.  The rough
  (inclination not removed) locations and masses of planets estimated
  to be present are shown at the bottom of the middle panel as yellow
  circles.  The circles on the left side of the figure (white/black)
  represent the angular resolution of the data.}
\label{fig:hd163}
\end{figure}

{\bf Detection of young exoplanets with molecules.} One new frontier
is the use of ALMA to infer the presence of hidden planets, first
using dust emission \citep{Andrews20} followed by
constraints from CO emission showing deep gas cavities and gaps
\citep{vanderMarel16}.  Even more recently, the presence of
hidden planets based on kinematical information has been gleaned from
ALMA CO emission maps \citep[][and other work by their
teams]{Teague18,Pinte18}. Fig.~\ref{fig:hd163} provides an example of
the overall emission (dust and CO) along with the exquisite measured
velocity field.  This topic is very much in its infancy but has
promise to potentially isolate the locations of hidden planets, set
unique constraints on the H$_2$ gas pressure, and reveal the presence
of ordered gas motions.

A central aspect of this work is the close collaboration with
dynamical theorists. Indeed, the wealth of dust substructures in disks
points to a larger number of giant exoplanets than found by direct
imaging at 10--100 au distance, suggesting a so-far unseen population
of sub-Jupiter mass planets that needs independent confirmation
\citep{Zhang18,Lodato19}.  The kinematical work is pure molecular
astrophysics, but has interesting astrochemical links. Different
molecular tracers probe different layers and one might use this
information to study the overall physics and chemistry of material
directly related to forming planets.

{\bf Exoplanet atmospheres.} The chemical composition and origin of
exoplanet atmospheres, from Super-Earths to mini-Neptunes and Jovian
planets, is clearly a new frontier for Astrochemistry
\citep{Madhusudhan19} and discussed in other chapters in this book.
One of the ultimate goals is to link the planetary atmosphere
composition with its formation history in the natal protoplanetary
disk. On the one hand, the changing C/O ratio with disk radius could
provide such a probe of the formation location
(Fig.~\ref{fig:oberg}). However, the route from disk gas+dust to a
mature planet is long and involves many steps, each of them with
significant uncertainties \citep[e.g.,][]{Mordasini16,
  Cridland19}. For example, are the heavy elements (i.e., other than H
and He) in a giant planet atmosphere accreted mostly from the gas or
delivered by icy pebbles? How does the migration history affect the
outcome?  How about dust traps? It is also clear that individual
molecule abundances will be fully reset in giant planet atmospheres,
preserving only the overall C/O, C/N, O/H etc. abundance ratios. And
even those could be affected if part of the atmosphere material is
cycled to the planetary core.

{\bf Rocky exoplanets, lessons from Earth.} The atmospheres of rocky
terrestrial planets may have an even more complicated history. Well
outside the water snowline, the planets are built up largely from
planetesimals that are roughly half rock and half ice. When these
planets move inward, water becomes liquid, resulting in ocean planets
or water worlds. Inside the snowline, the planets are usually thought
to be very dry. Computing the atmospheric composition of terrestrial
exoplanets is significantly more complex than that of giant exoplanets
\citep{Kaltenegger17}.  In this light, a central conundrum lies in the
Earth's carbon content which is orders of magnitude below the carbon
content seen in cometary material and the Sun.
Overall, this hints at loss of the main carriers of carbon (and also
nitrogen) in Earth's planetary building blocks in the inner few au of
the presolar nebula, and in disks more generally. It also requires
consideration of many additional processes that are operative during
each stage of the formative process of terrestrial worlds
\citep[e.g.][]{Bergin15,Gail17}, followed by billions of years of
evolution on a living active planet \citep{Foley19}.

A secondary atmosphere can however be formed through impacts of (icy)
planetesimals in the late formation stages, including by comets such as
67P/C-G which can deliver water and organics to the young planet.
Given the architecture of the solar system it is not clear if
significant mass from comet-forming zones can be brought to Earth
\citep{vanDishoeck14ppvi}. However, the giant planets could have migrated
and noble gas analyses of 67P are consistent with some cometary supply
to the young Earth including most of the organic material, even if not
the water \citep{Marty17}. The survival of these organics depends on
planetesimal size and impact speed, and whether the volatile material
can perhaps be shielded by a protective layer on the parent body. If
they do, there would indeed be a direct link between interstellar
molecules and the building blocks for life on new planets.

Key references for this section include
\cite{Disk20,Mordasini16,Madhusudhan19,Bergin15}.

\section{Opportunities} 

There are several frontiers in front of us.  On short timescales ALMA
remains a pioneering instrument and JWST is on the horizon.  The
combination of these two observatories will be quite powerful as we
describe below.

\begin{itemize}

\item ALMA will be able to survey hundreds of young disks in the
  embedded phase for their chemical composition and thus search for
  similarities and differences in the chemistry prior to planet
  formation across stellar mass and age range. ALMA will also reveal
  whether substructures in young disks are common or not, which could
  point to locations where planets are typically formed.

\item ALMA has inferred the presence of volatile depletion in disk
  systems beyond the CO snowline, and in a few instances interior to
  the CO snowline.  JWST will be able to probe the inner disk
  composition (within a few AU, including O and C budgets) in the
  upper disk atmosphere as well as that of ices in the intermediate
  layer for a few near edge-on disks. Will ALMA and JWST tell the same
  story?

\item In a handful of systems the stellar abundances, and that of
  accreted/ejected material, provide a crucial link to explore the
  composition of gas in the inner disk that reaches the star. Linking
  these studies to ALMA/JWST analyses may provide unique information
  on the composition of hidden material.

\item Models of grain growth to pebble size and their radial drift are
  becoming increasingly sophisticated and are being coupled with
  chemistry.  Since pebble accretion is believed to be a primary
  mode of planetary growth, these models (combined with observations)
  are a needed step.

\item ALMA has just begun to explore the kinematical landscape of planet formation.  A clear question is whether any planets inferred to be present by molecular kinematics, or emission structures, are confirmed by ground-based or space-based direct detection.  Within this landscape the link between chemistry and 3D kinematics presents an inviting field to explore the composition of gas that is being supplied from the surface to the planet-forming zone.

\item High spectral resolution infrared spectrometers on future
  extremely large telescopes, such as ELT/METIS, will be able to
  spatially resolve the chemistry and kinematics of warm gas down to a
  few au and also image, in tandem with ALMA, gas and dust in
  circumplanetary disks. VLTI/Gravity(+) provides new opportunities to
  obtain spatially resolved spectra of young exoplanets and their
  disks on even smaller scales.

\end{itemize}

  \bibliographystyle{aa}
\setlength{\bibsep}{0.0pt}

\end{document}